\documentclass[pra,twocolumn,showpacs,floatfix]{revtex4}

\usepackage{amsmath}
\usepackage{times}
\usepackage{txfonts}
\usepackage{graphicx}
\usepackage{color}
\usepackage{float}
\usepackage{soul}
\usepackage{tabularx}
\usepackage{booktabs}
\usepackage{bm}
\usepackage[colorlinks,linkcolor=blue,citecolor=blue,urlcolor=blue]{hyperref}
\graphicspath{Figures/}
\hypersetup{urlcolor=blue, colorlinks=true}

\begin{document}

\newcolumntype{L}[1]{>{\raggedright\arraybackslash}p{#1}}
\newcolumntype{C}[1]{>{\centering\arraybackslash}p{#1}}
\newcolumntype{R}[1]{>{\raggedleft\arraybackslash}p{#1}}

\title{Effects of self- and cross-phase modulation on the spontaneous symmetry breaking of light in ring resonators} 

\author{Lewis \surname{Hill$^\dagger$}}
\email{lewis.hill@strath.ac.uk}
\author{Gian-Luca \surname{Oppo}} 
\affiliation{Department of Physics, 
University of Strathclyde, 107 Rottenrow, Glasgow G4 0NG, Scotland, UK, and SUPA\\$^\dagger$ also with National Physical Laboratory, Hampton Road, Teddington, TW11 0LW}
\author{Michael T. M. \surname{Woodley$^\ddagger$}}
\author{Pascal \surname{Del'Haye}}
\affiliation{National Physical Laboratory, Hampton Road, Teddington, TW11 0LW, UK\\$^\ddagger$ also with Heriot-Watt University, Edinburgh, EH14 4AS, UK}  

\begin{abstract}
We describe spontaneous symmetry breaking in the powers of two optical modes coupled into a ring resonator, using a pair of coupled Lorentzian equations, featuring tunable self- and cross-phase modulation terms. We investigate a wide variety of nonlinear materials by changing the ratio of the self- and cross-phase interaction coefficients. Static and dynamic effects range from the number and stability of stationary states to the onset and nature of oscillations. Minimal conditions to observe symmetry breaking are provided in terms of the ratio of the self-/cross-phase coefficients, detuning, and input power. Different ratios of the nonlinear coefficients also influence the dynamical regime, where they can induce or suppress bifurcations and oscillations. A generalised description on this kind is useful for the development of all-optical components, such as isolators and oscillators, constructed from a wide variety of optical media in ring resonators.
\end{abstract}


\maketitle

\section{Introduction}
Originally proposed in 1987, the Lugiato-Lefever equation (LLE) \cite{lugiato1987spatial} has been used to model a variety of nonlinear optical systems \cite{chembo2017theory}. One of the equation's major successes has been in describing light propagating in fibre loops and micro-ring resonators featuring Kerr media -- materials in which the refractive index is modified by the intensity of the propagating light. While the original model described spatio-temporal dynamics in the plane transverse to the direction of propagation, a later model for purely temporal effects \cite{haelterman1992dissipative} has been demonstrated to be mathematically equivalent \cite{lugiato2015nonlinear,castelli2017lle}.

Coupled LLEs have been used to describe normalised left- and right-circularly-polarised field envelopes, $E{\pm}$, in Fabry-Perot or ring cavities \cite{geddes1994polarisation}. This system of two coupled LLEs is given by
\begin{equation}\label{EQ:1}
\frac{\partial E_{\pm}}{\partial t}= E_{In} - E_{\pm}-i\eta\theta E_{\pm}+ia\nabla^2E_{\pm} 
+i\eta\left(A|E_{\pm}|^2+B|E_{\mp}|^2\right)E_{\pm}
\end{equation}
where $\theta$ denotes the cavity detuning -- the difference between the input field's frequency and the closest cavity resonance frequency, $a$ describes the transverse diffraction strength, $E_{\mathrm{In}}$ is the input pump envelope, $\eta=\pm1$ indicates either a self-focusing, $+1$, or self-defocusing medium, $-1$, respectively, and $\nabla^2$ is the transverse Laplacian. The one-dimensional case of Eq.~\eqref{EQ:1}, with diffraction replaced by dispersion, describes the propagation of two optical field components in ring resonators. The coupling constants $A$ and $B$ are related to the third-order nonlinear susceptibility tensor, $\chi^{(3)}$, and describe the strengths of self- and cross-phase modulation, respectively -- the former is the change in refractive index induced by one optical mode on itself, and the latter is the change induced by the other mode. The values of these two coefficients are of great interest because their relative magnitudes and signs vary across a large number of different experimental configurations. These situations include light propagating through dielectrics, optical fibres, Kerr liquids (such as carbon disulphide, benzene, toluene, and certain liquid crystals), engineered structures such as periodically-poled lithium niobate, as well as experiments featuring atomic vapours. We provide here a comprehensive theory of spontaneous symmetry breaking in the intensity of two distinct optical modes, across a variety of different experimental contexts, by considering variations of the ratio $B/A$ -- the central parameter of the investigations presented here. 

By restricting the solution set of $E_{\pm}$ in Eq.~\eqref{EQ:1} to being both stationary and homogeneous, and then multiplying each element by its complex conjugate, one obtains

\begin{equation}\label{EQ:2}
|E_{\pm}|^2=\frac{E_{In}^2}{1+\left(\theta-A|E_{\pm}|^2-B|E_{\mp}|^2\right)^2}.
\end{equation}

\noindent This particular solution corresponds to two coupled Lorentzian equations -- mathematically identical to those that describe two normalised counter-propagating stationary fields in ring resonators \cite{kaplan1981enhancement,kaplan1982directionally,del2017symmetry,del2018microresonator,wright1985theory,woodley2018universal}. Of course, mathematical equivalence does not necessarily imply physical equivalence. In the counter-propagating case, $E_{\pm}$ are the two counter-propagating field envelopes and the coupling constants $A$ and $B$ now depend on the formation of an index grating generated by the two fields, rather than on $\chi^{(3)}$ as with the polarisation equations \cite{otsuka1983nonlinear, firth1985diffusion, firth1988transverse, firth1990transverse}.

\clearpage

For ease of notation, we set $|E_+|^2=P_1$, $|E_-|^2=P_2$ and $E_{\mathrm{In}}^2=I$, such that Eq.~\eqref{EQ:2} may be expressed as

\begin{equation}\label{EQ:3}
P_{1,2}=\frac{I}{1+\left(\theta-AP_{1,2} - B P_{2,1}\right)^2}.
\end{equation}

\noindent Equation \eqref{EQ:3} can be understood as the homogeneous stationary solution set of any system described by two coupled LLEs, such that many of the subsequent results of this paper can be applied not only to both the counter-propagating and polarisation cases, but to other physical systems, too.

One fascinating phenomenon that arises from a system of two coupled Lorentzian equations, such as Eq.~\eqref{EQ:2}, is spontaneous symmetry breaking \citep{kaplan1982directionally,wright1985theory}. We first extend the investigation of the onset of symmetry breaking in ring resonators to a variable ratio of the self- and cross-phase modulation terms, $B/A$, in Section II. We then identify the steady-state characteristics of the symmetry breaking for variable $B/A$ in Section III. Sections IV and V are devoted to an analytical stability analysis and dynamical behaviour via numerical integration, respectively. In the latter case, we ascertain how varying the cross-coupling strength between the two fields alters the temporal instability of the system, thereby encouraging or suppressing deterministic chaos. Our conclusions are summarised in Section VI.

We note that symmetry breaking phenomena have a wide range of applications in nonlinear optics: enhancing the Sagnac effect \cite{kaplan1981enhancement,wright1985theory}; realising isolators, circulators \cite{del2018microresonator}, and all-optical oscillators \cite{woodley2018universal} (for use in integrated photonic circuits); and the development of enhanced near-field detectors \cite{wang2015nonlinear}. A possible area of further application is the generation of temporal cavity solitons (TCS). TCS are known to be of great future utility \cite{grelu2015nonlinear} in areas such as data storage \cite{gilles2017polarization, haelterman1994polarisation, haelterman1995colour, wabnitz2009cross} and in the generation of frequency combs \cite{leo2010temporal,coen2013modeling,leo2013dynamics,coen2013universal,herr2014temporal,
parra2014dynamics}. There is enormous current interest in extending the range and realisation of TCS due to their diverse utility in fields such as precision metrology, gas sensing, arbitrary optical waveform generation, and telecommunications \cite{del2007optical,kippenberg2011microresonator,papp2014microresonator,okawachi2011octave,
ferdous2011spectral,herr2012universal,pfeifle2014coherent}.

\section{Spontaneous Symmetry Breaking}
\label{Symmetry}
Spontaneous symmetry breaking of two modes in an optical ring resonator manifests itself as unequal coupling of the two input powers into the resonator. Consequently, we will refer to spontaneous symmetry breaking of the `coupled powers'. This was first predicted theoretically in Ref.~\cite{kaplan1982directionally}, and has since been experimentally observed in Refs.~\citep{del2017symmetry,del2018microresonator} for counter-propagating fields, whilst the polarisation case is discussed in Refs.~\citep{geddes1994polarisation, gallego2000, Copie2019}, and in an experimental context in Ref.~\cite{Fatome:18}. 

Spontaneous symmetry breaking in the coupled Lorentzian system can be visualised in a number of ways. One way is to eliminate the explicit dependence on the pump power, $I$, by rearranging Eq.~\eqref{EQ:3} such that the two expressions are each made equal to $I$. They may then be solved simultaneously as

\begin{equation}\label{EQ:4}
P_{1}\left[1+\left(\theta-AP_{1}-BP_{2}\right)^2\right]=P_{2}\left[1+\left(\theta-AP_{2}-BP_{1}\right)^2\right]\;.
\end{equation}

\begin{figure}[h]
\centering
\includegraphics[width=0.48\textwidth]{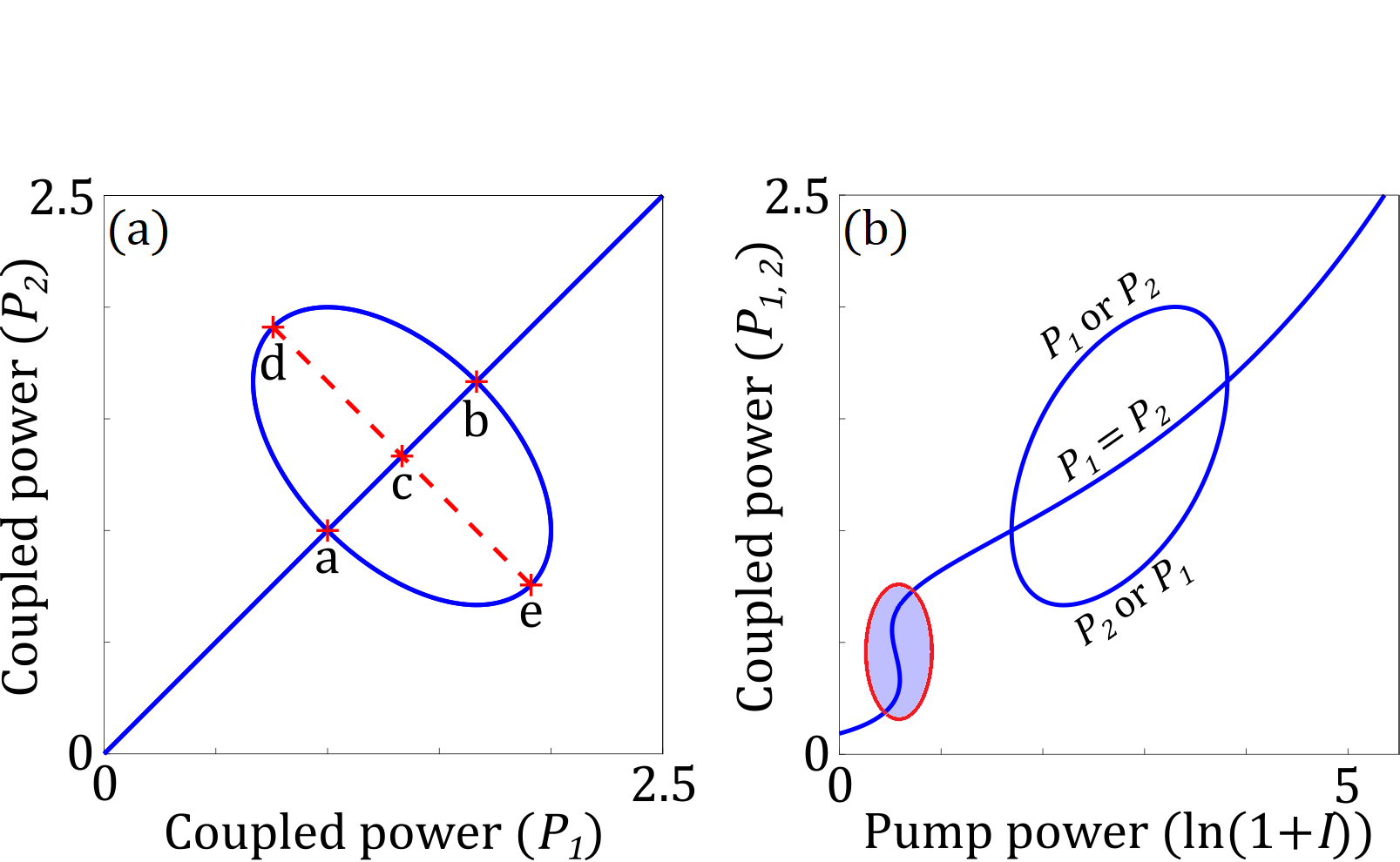}
\caption{Different graphical representations of spontaneous symmetry breaking when scanning pump power, $I$, shown here for $A=1$, $B=2$ and $\theta=2$. (a) Coupled powers are plotted against each other, Eq.~\eqref{EQ:4}. The points a and b indicate the opening and closing of the symmetry-broken bubble given by Eq.~\eqref{EQ:13}.  Point c is the point at which symmetry breaking occurs at the detuning limit, see Eq.~\eqref{EQ:14}. The maximum coupled power difference can be found at points d and e, see Eq.~\eqref{EQ:16}. (b) The coupled powers are plotted against input power. Note the visible presence of the `symmetric bistability' (the highlighted S-shaped curve), but also the loss of some symmetries seen in (a).}
\label{fig:1}
\end{figure}

This solution is plotted in Fig.~\ref{fig:1}(a), and corresponds to a `scan' with respect to the pump power, $I$, shown in Fig.~\ref{fig:1}(b). The `symmetric' solution line features as a simple $P_{1}=P_{2}$ relationship, and the spontaneous emergence of the symmetry-broken solution line is characterised by an ellipse. On the symmetric solution line, both field envelopes exhibit equal intensities, which clearly breaks down on the symmetry-broken curve. The point at which symmetry-broken solutions become possible is known as the `symmetry breaking bifurcation point', whereas the point at which they disappear is the `inverse bifurcation point'.

It has been shown that, in the case of $A=1$, $B=2$, the symmetric solution line between the bifurcation points is unstable, and so, if the system is subject to a perturbation, such as noise, it will evolve towards the stable symmetry-broken solution line \citep{kaplan1982directionally}. This is an extremely useful result, since it means that the two observed field envelopes will no longer circulate with equal intensity -- one field envelope will become dominant, whilst the other is quenched. This behaviour is central to the applications mentioned previously.

Fig.~\ref{fig:1}(a) is the counterpart of Fig.~\ref{fig:1}(b), originally reported in Ref.~\cite{kaplan1982directionally}). In different ways, they both illustrate the symmetry breaking by scanning the input power. An informative advantage of Fig.~\ref{fig:1}(b) comes from its ability to show the `symmetric bistability' -- highlighted by a red ring. This region is present in Fig.~\ref{fig:1}(a), but is hidden within the symmetric solution line. The advantage of Fig.~\ref{fig:1}(a), however, comes from its additional symmetry, which can allow for mathematical simplifications in the derivations of later results.

\clearpage

\begin{figure}[h]
\centering
\includegraphics[width=0.48\textwidth]{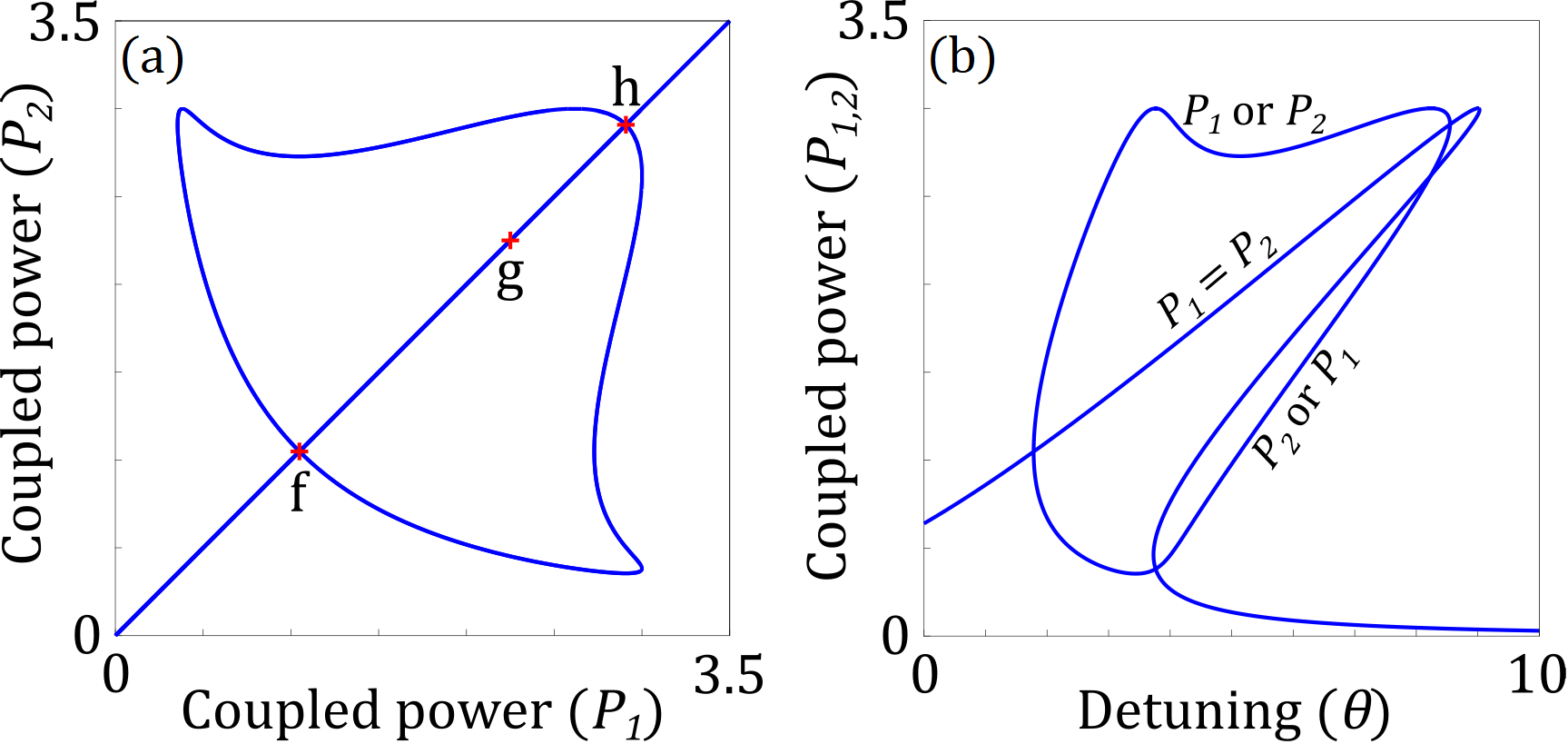}
\caption{Different graphical representations of spontaneous symmetry breaking when scanning detuning, $\theta$, shown here for $A=1$, $B=2$ and $I=3$. (a) Coupled powers are plotted against each other, Eq.~\eqref{EQ:5}. The points f and h indicate the symmetry breaking bifurcation pair for a detuning scan -- see Eq.~\eqref{EQ:18}. At point g, the bubble emerges at the intensity limit -- see Eq.~\eqref{EQ:17}. (b) Coupled envelope powers are plotted against detuning, $\theta$. Note again the loss of symmetry between (a) and (b)}
\label{fig:2}
\end{figure}

It is also possible to observe symmetry breaking when scanning the cavity detuning rather than the pump power. This can be done by employing a similar method to above -- by rearranging Eq.~\eqref{EQ:3} such that the two expressions are in terms of $\theta$; they can again be solved simultaneously, eliminating $\theta$,

\begin{equation}\label{EQ:5}
AP_{1}+BP_{2}\pm\sqrt{\frac{I}{P_{1}}-1}=AP_{2}+BP_{1}\pm\sqrt{\frac{I}{P_{2}}-1}\;,
\end{equation}

\noindent where each $\pm$ is independent of the other. This solution set is plotted in Fig.~\ref{fig:2}(a), along with its analogous graph, \ref{fig:2}(b) , reported in \cite{woodley2018universal}.

Figs.~\ref{fig:1}(b) and \ref{fig:2}(b) can be obtained by rearranging one of the coupled Lorentzian equations such that it is equal to one of the variables $P_{1,2}$, and substituting this into the second of the Lorentzian equations.

It is possible, for all graphs contained within Figs.~\ref{fig:1} and \ref{fig:2}, to isolate the symmetric and symmetry-broken solution curves using the following methods:

\textit{Symmetric solution line} -- set $P_{1}=P_{2}$ in each of the coupled equations, then simplify. Retain both $P_{1}$ and $P_{2}$ as separate variables, however, to allow for simultaneous plotting.

\textit{Symmetry-broken solution line} -- take the full equation describing the solution set and divide by the equation describing the symmetric solution set, then simplify.

By studying each component individually, the mathematical complexities of an analysis can in some cases be drastically reduced.

Many of the applications described previously require careful predictions about the characteristics of the symmetry broken region. Some of these characteristics, such as the minimum detuning required for symmetric bistability, or the possibility for symmetry-broken solutions, have been reported for specific values of $A$ and $B$: $A=1$, $B=2$ in the case of Ref.~\cite{kaplan1982directionally}. A larger, but finite, range is analysed in Ref.~\cite{martin2013codimension,martin2010homogeneous}, but there appears to be no general analysis spanning all values of $A,B\in\mathbb{R}$. We present this general analysis here along with useful results that are pertinent to the applications mentioned above. Firstly, however, a more immediate question presents itself: which values of $A$ and $B$ are physically feasible? The answer, of course, depends on the experimental situation.

In the case of two coupled Lorentzian equations describing two counter-propagating fields, the symmetry breaking is a result of the formation of an index grating in the medium due to the standing wave interference pattern that forms \cite{otsuka1983nonlinear, firth1985diffusion, firth1988transverse, firth1990transverse}. In this case, the values that $A$ and $B$ can take are given by $A=1$, $B=1+h$, where $0\leq h\leq1$, depending on the medium's ability to `wash out' the grating via, for example, diffusion, in the case of a gas or liquid. In a medium with no diffusive effects, $h=1$, whilst for a highly mobile Kerr medium, such as a gas, $h\rightarrow0$.

The polarisation case has far greater variation in the possible values that the coupling constants can take. In this case, $A$ and $B$ are related to the third-order nonlinear susceptibility tensor, $\chi^{(3)}$, by
\begin{equation}\label{EQ:6}
A=\frac{\chi_{1122}^{(3)}+\chi_{1212}^{(3)}}{\chi_{1111}^{(3)}},\;\;\;\;\;B=\frac{\chi_{1122}^{(3)}+\chi_{1212}^{(3)}+2\chi_{1221}^{(3)}}{\chi_{1111}^{(3)}}\;,
\end{equation}

\noindent with the constraint that $A+B=2$ for an isotropic medium \cite{geddes1994polarisation}. The other cases are: a nonresonant electronic response, $A=2/3$, $B=4/3$; liquids or molecular orientation, $A=1/4$, $B=7/4$; and electrostriction, $A=1$, $B=1$ \cite{boyd2003nonlinear}. Deviating momentarily from Kerr media, atomic vapours are likely to show phenomena offering a wide range of possible magnitudes of $A$ and $B$  \cite{geddes1994polarisation,geddes1994patterns}, experimentally shown in Ref.~\cite{burgin2005femtosecond}. These atomic vapours could be used, for example, in hollow fibres. In addition, we believe that it may be possible to access negative values of $B/A$ by appropriate engineering of a ring resonator exhibiting an effective $\chi^{(3)}$ nonlinearity, such as periodically-poled lithium niobate (PPLN) \cite{das2006modulation,miyata2009phase, balachninaite2000self}. These $A$ and $B$ values are summarised in Table.~\ref{table}.

The values of these coupling constants may not be purely limited to those suggested here. For example, nonlinear thermal effects \cite{Carmon2004} act to rescale $A$ and $B$ by equal amounts -- i.e., they are symmetric effects. The following analysis can be applied to any system described by coupled LLEs or Lorentzian equations of the forms given by Eq.~\eqref{EQ:1} and Eq.~\eqref{EQ:3}, respectively, such as in Ref.~\cite{martin2010homogeneous}, where both electric and magnetic nonlinearities are modelled.

\begin{table}[H]
\centering
\begin{tabular}{lC{1.8cm}C{1.8cm}}
\hline
\textbf{Counter-propagating fields} & \textbf{A} & \textbf{B} \\
\hline
Solids (without diffusion) & 1 & 2 \\

General diffusive effects & 1 & $1+h$ $(0<h\leq 1)$ \\

Gases (high rates of diffusion) & 1 & $\rightarrow$ 1 \\ 
\hline
\textbf{Polarisation effects} & & \\
\hline
Isotropic media & \multicolumn{2}{c}{$A + B = 2$}\\
Non-resonant electronic response & 2/3 & 4/3\\
Liquids, or molecular orientation & 1/4 & 7/4\\

Electrostriction & 1 & 1\\

$\chi^{(2)}$ media with effective $\chi^{(3)}$ & \multicolumn{2}{c}{Potentially negative $B/A$} \\

Atomic vapours & \multicolumn{2}{c}{Wide range of $B/A$} \\
\hline      
\end{tabular}
\caption{A selection of different experimentally-accessible values of $A$ and $B$.}
\label{table}
\end{table}

\clearpage

\section{Changing the relative strengths of self- and cross-phase modulation}
\label{GenRes}

The first generalised result observed here is the region of optical bistability for symmetric solutions, previously seen highlighted in Fig.~\ref{fig:1}(b) with a red ring. The symmetric solution line in the circulating powers vs. input power diagrams is given by
\begin{equation}\label{EQ:7}
I=P_{1,2}\left\{1+\left[\theta-\left(A+B\right)P_{1,2}\right]^2\right\}\;,
\end{equation}

\noindent The bistable region is found to be bounded by the following:
\begin{equation}\label{EQ:8}
P_{1,2}=\frac{2\theta\pm\sqrt{\theta^2-3}}{3(A+B)}\;,
\end{equation}

\noindent where $dI/d(P_{1,2})=0$. This reveals that there is a limiting detuning value for symmetric optical bistability of $\theta=\sqrt{3}$ that is independent of the values of the coupling constants. The coupled powers themselves, however, are dependent on the coupling constants. Inserting Eq.~\eqref{EQ:8} into Eq.~\eqref{EQ:7} gives the limits on the input power, between which lies the region of symmetric bistability,
\begin{equation}\label{EQ:9}
I=-\frac{2\left(2\theta\pm\sqrt{\theta^2-3}\right)\left(-\theta^2\pm\theta\sqrt{\theta^2-3}-3\right)}{27(A+B)}\;.
\end{equation}

\begin{figure}[h]
\centering
\includegraphics[width=0.25\textwidth]{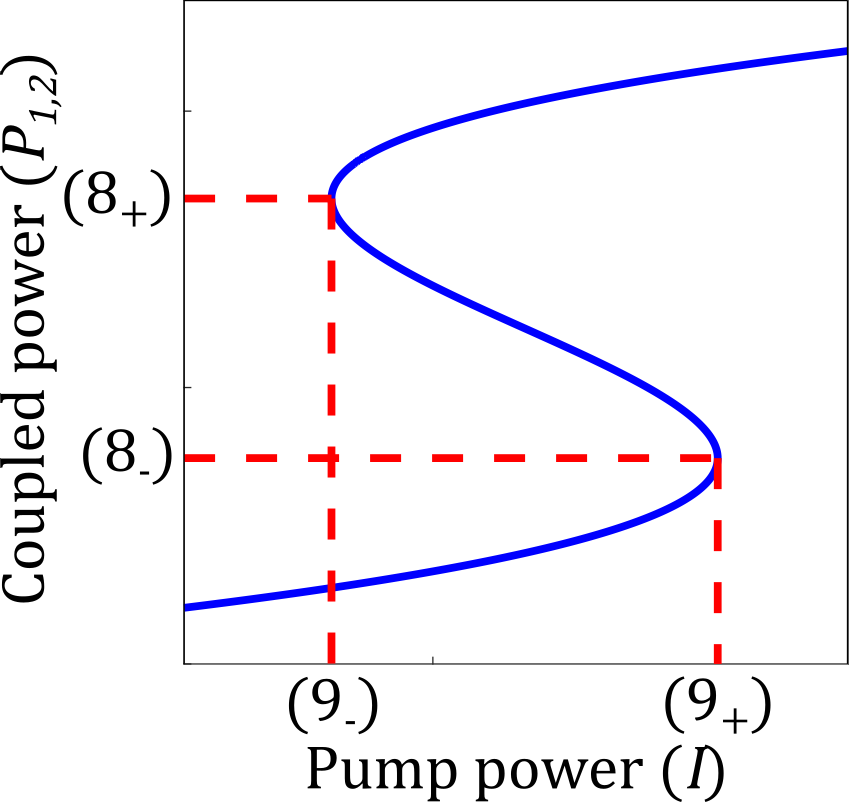}
\caption{Example optical bistability of Eq.~\eqref{EQ:7}, with limits calculated via Eqs.~\eqref{EQ:8} and \eqref{EQ:9}.}
\label{fig:3}
\end{figure}

\noindent These pump power limits are also dependent on $A$ and $B$, with higher values of $A+B$ leading to a lower value of required input power. Note that in Eq.~\eqref{EQ:9}, a choice of one $\pm$ sign enforces the same choice on the other. A graphical example of these results is given in Fig.~\ref{fig:3}.

The characteristics of the symmetry-broken region are most easily analysed by examining the symmetry-broken part of Eq.~\eqref{EQ:4}, which is given by
\begin{equation}\label{EQ:10}
\left[\theta-A\left(P_{1}+P_{2}\right)\right]^2-P_{1}P_{2}\left(B-A\right)^2=-1\;.
\end{equation}

\noindent Deriving $dP_{2}/dP_{1}$ in Eq.~\eqref{EQ:10} and imposing the condition that $dP_{2}/dP_{1}=-1$, the detuning limit for symmetry-broken solutions can be ascertained. For cavity detunings below this limit, the symmetry broken region will never emerge, for any pump power. This detuning limit is given by
\begin{equation}\label{EQ:11}
|\theta| > \frac{\sqrt{\left(3-\frac{B}{A}\right)\left(1+\frac{B}{A}\right)}}{\left|\frac{B}{A}-1\right|}\;,
\end{equation}

\begin{figure}[h]
\centering
\includegraphics[width=0.475\textwidth]{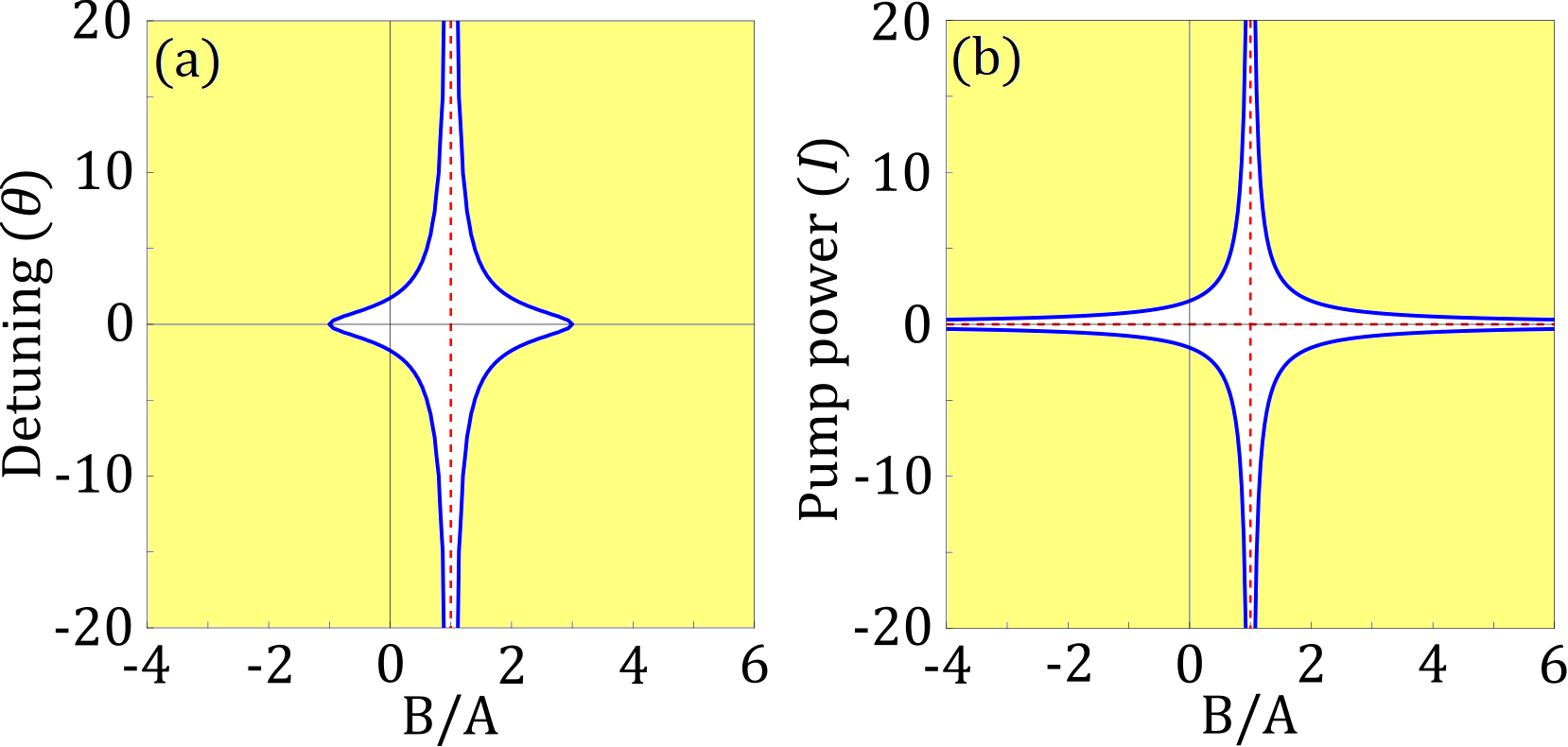}
\caption{(a) Minimum detuning required to observe symmetry breaking when changing the ratio of the coupling constants $B/A$. 
The yellow regions indicate where symmetry breaking is possible, with the blue lines indicating the limits where symmetry breaking becomes impossible. (b) Minimum input power required to observe symmetry breaking. Negative values for $I$ are included in b) for completeness, but are not physically attainable.}
\label{fig:4}
\end{figure}

\noindent As shown in Fig.~\ref{fig:4}, this equation reveals two important points of interest. The first one is that, for a unity ratio between the two coupling constants, symmetry breaking is never possible, since $\theta_{min}$ diverges to $\infty$. The second interesting point is that, for $B/A>3$ or $B/A<-1$, symmetry breaking is attainable for all detuning values, even $\theta=0$, for pump powers above given thresholds.

If one instead wishes to minimise the pump power requirement, the input limit is given by

\begin{equation}\label{EQ:12}
|I| > \frac{\frac{8}{9}\sqrt{3}}{\left|\frac{B}{A}-1\right|}\;.
\end{equation}

\noindent Again, below this limit, symmetry breaking is not possible for any range of cavity detunings. Unlike with the detuning limit, this power limit only falls to $0$ as $B/A$ tends to $\infty$.

The analysis for $dP_{2}/dP_{1}=-1$ also reveals the coupled powers at which the symmetry breaking bifurcation points are located. These points, where the symmetry-broken region opens/closes, are given by
\begin{equation}\label{EQ:13}
P_{1,2}=C\pm\frac{D}{\left(3A-B\right)\left(A+B\right)}\;,
\end{equation}

\noindent where
\begin{equation}\label{EQ:14}
C=\frac{2A\theta}{\left(3A-B\right)\left(A+B\right)}\;,
\end{equation}

\noindent and
\begin{equation}\label{EQ:15}
D=\sqrt{-3A^2+\theta^2\left(A-B\right)^2-2AB+B^2}\;.
\end{equation}

\noindent Further analysis of the symmetry-broken solution curve reveals general results that are of importance for the optimisation of the formation of isolators for integrated photonic circuits, such as in Ref.~\cite{del2018microresonator}. For such applications, one mode must be suppressed as much as possible, whilst the other mode is maximised. By obeying the constraint $dP_{2}/dP_{1}=1$, one can obtain the coupled powers of the greatest possible difference
\begin{equation}\label{EQ:16}
P_{1,2}=C\pm\frac{1}{B-A}\frac{D}{\sqrt{\left(3A-B\right)\left(A+B\right)}}\;.
\end{equation}

\noindent These special points are summarised in Fig.~\ref{fig:1}(a) (points a, b, c, d, e) and  Fig.~\ref{fig:2}(a) (points f, g, h) with the input power required to reach each point given by substituting the appropriate equations into Eq.~\eqref{EQ:3}. 

We observe that Eq.~\eqref{EQ:13} identifies a `bursting' ratio between the constants, above which the symmetry-broken region opens, but never closes. Consequently, isolators based on this principle would have no upper limit of operational power (above which they would return to symmetric solutions). This bursting ratio, above which the symmetry-broken solution line forms a parabola rather than an ellipse, is given by $B/A>3$ or $B/A <-1$.

Turning attention to Eq.~\eqref{EQ:5}, some key points of the detuning scans can be identified. At the power limit, Eq.~\eqref{EQ:12}, the symmetry-broken region emerges at
\begin{equation}\label{EQ:17}
P_{1,2}=\frac{3}{4}I\;,
\end{equation}

\noindent while the symmetry breaking bifurcation point pair is given by solving the real roots to the quartic equation
\begin{equation}\label{EQ:18}
P_{1,2}^4-IP_{1,2}^3+\left[\frac{I}{2(A-B)}\right]^2=0\;.
\end{equation}

\noindent The detuning requirements to observe these points can then be obtained by substituting the appropriate equations into Eq.~(\ref{EQ:3}).

\begin{figure}[h]
\centering
\includegraphics[width=0.48\textwidth]{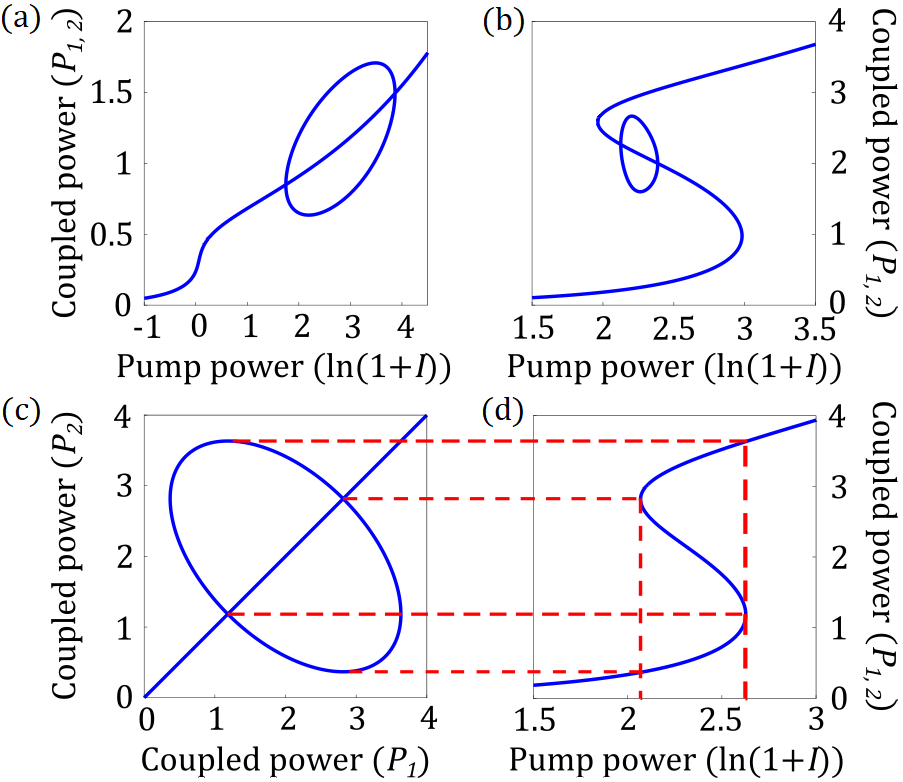}
\caption{(a), (b) Plots of the input power scans for $B/A=2.2$, $\theta=1.5$ and $B/A=0.5$, $\theta=4$ respectively. (c), (d) show the special case of $B=0$, $A\in\mathbb{R}$, with $A=1$, $\theta=3$. (c) shows the possibility still for symmetry-broken solutions, while (d) shows how, in this special case, their origin is due to the bistable region of the Lorentzian equation.}
\label{fig:5}
\end{figure}

In closing this section, we note that the value of $B/A$ also affects where the symmetry-broken solution line appears with respect to the bistable symmetric solution line. It is known that, for $A=1$ and $B=2$, the symmetry-broken `bubble' appears on the upper branch of the bistable symmetric solution line for graphs like that of Fig.~\ref{fig:1}(b) \citep{kaplan1982directionally}. This is because, for this $B/A$ ratio, Eq.~(\ref{EQ:11}) dictates that symmetry-broken solutions are only possible for $\theta\geq\sqrt{3}$, with $\theta=\sqrt{3}$ being the condition where optical bistability emerges. This holds true for any $1<B/A\leq2$. Above ratios of $2$, the minimum detuning for symmetry breaking is below that for optical bistability, meaning that it is now possible to observe the symmetry-broken solutions without bistability, Fig.~\ref{fig:5}(a). More interesting is the region $B/A<1$. For $0<B/A<1$, symmetry breaking is again only possible for detunings above the $\sqrt{3}$ value for optical bistability, but now the symmetry-broken bubble appears on the middle branch of the bistable region, as shown in Fig.~\ref{fig:5}(b). Progressing further, for $B/A<0$, it is once again possible to observe the symmetry broken solutions for detunings lower than the minimum required for symmetric solution line optical bistability.

The only ratio not covered by the regions described above is the special case of $B/A=0$. Plotting in the style of Fig.~\ref{fig:1}(a) for $B=0$ and generic values of $A$, symmetry broken solutions are, interestingly, still possible, as shown in Fig.~\ref{fig:5}(c) for $A=1$. The value of $A$ changes only the required input powers. This explains the continuous nature of all equations described previously, and Fig.~\ref{fig:4}, about $B/A=0$. This symmetry breaking is not due to any cross-talk of the coupled powers. Rather, it is due to the arbitrary constraint imposed that both $\theta$ and $I$ are equal for both equations. This results in the two, now uncoupled, Lorentzian equations being identical, Fig.~\ref{fig:5}d. The symmetry broken solutions arise physically from the possibility of one field being on the top branch of the optical bistability while, simultaneously, the other is on the bottom, or vice versa.

\section{Generalised Stability Analysis}
\label{GenStab}

In the same spirit as in Ref.~\cite{woodley2018universal}, we recognise that Eq.~\eqref{EQ:3} is the steady state of the following time-dependent system:

\begin{equation}\label{EQ:19}
\frac{\partial E_{\pm}}{\partial t}=E_{\mathrm{in}}-[1+i(\theta-A|E_{\pm}|^{2}-B|E_{\mp}|^{2})]E_{\pm}.
\end{equation}

\noindent Following the procedure set out in Ref.~\cite{woodley2018universal}, we add small perturbations to the steady state solution, calculate the eigenvalues of the (Jacobian) matrix that results, and assess the stability of this system. 

\clearpage

The eigenvalues of the linear stability of Eq.~\eqref{EQ:19} have the same form as that provided in Ref.~\cite{woodley2018universal}:

\begin{equation}\label{EQ:20}
\lambda=-1\pm\sqrt{\frac{-\alpha_{1}\beta_{1}-\alpha_{2}\beta_{2}\pm S}{2}},
\end{equation}

\noindent with

\begin{equation}\label{EQ:21}
S=\sqrt{(\alpha_{1}\beta_{1}-\alpha_{2}\beta_{2})^{2}\!+4\alpha_{1}\alpha_{2}\gamma^{2}}
\end{equation}  

\noindent but the quantities $\alpha_{1,2}$, $\beta_{1,2}$ and $\gamma^{2}$ take on forms generalised to arbitrary self- and cross-phase modulation coefficients: $\alpha_{1,2}=\theta-AP_{1,2}-BP_{2,1}$, $\beta_{1,2}=\theta-3AP_{1,2}-BP_{2,1}$, and $\gamma^{2}=4B^{2}P_{1}P_{2}$. Note that in Eq.~\eqref{EQ:20} one $\pm$ choice enforces no restrictions on the other $\pm$, giving a total of four eigenvalues.

When examining these eigenvalues, the quantity $S$ plays an essential role in establishing the stability of the system. If $S$ is real, and the quantity under the square root in Eq.~\eqref{EQ:20} is negative for both $\pm S$, i.e, $S<\alpha_{1}\beta_{1}+\alpha_{2}\beta_{2}$, then all the eigenvalues are complex numbers with real part equal to $-1$, leading to full stability of the corresponding stationary states. On the other hand, if $S$ is real, and the quantity under the square root in Eq.~\eqref{EQ:20} is positive, then one real eigenvalue can be positive (the condition for non-oscillatory instability) if
\begin{equation} \label{EQ:22}
S > 2 + \alpha_1 \beta_1 + \alpha_2 \beta_2\;,
\end{equation}
with the maximum of two real eigenvalues being positive when
\begin{equation} \label{EQ:23}
S < - ( 2 + \alpha_1 \beta_1 + \alpha_2 \beta_2)\;
\end{equation}
is also satisfied. Note that this condition for a second unstable eigenvalue is only possible when $2+\alpha_1 \beta_1 + \alpha_2 \beta_2<0$.

Under the condition of $S$ being purely imaginary, the eigenvalues Eq. (\ref{EQ:20}) are complex with the real ($R$) and imaginary ($\Omega$) parts, corresponding to the growth rate and the angular frequency respectively, taking the following forms \cite{woodley2018universal}:
\begin{equation}\label{EQ:24}
R=-1\pm\sqrt{\frac{1}{2}\sqrt{\alpha_{1}\alpha_{2}(\beta_{1}\beta_{2}-\gamma^{2})}-\frac{1}{4}(\alpha_{1}\beta_{1}+\alpha_{2}\beta_{2})} \; ,
\end{equation}

\begin{equation}\label{EQ:25}
\Omega=\pm\sqrt{\frac{1}{2}\sqrt{\alpha_{1}\alpha_{2}(\beta_{1}\beta_{2}-\gamma^{2})}+\frac{1}{4}(\alpha_{1}\beta_{1}+\alpha_{2}\beta_{2})} \; .
\end{equation}
The instabilities are then obtained by finding the conditions for which $R>0$, and correspond to 
\begin{equation}\label{EQ:26}
|S^2|> 8(2+\alpha_1 \beta_1 + \alpha_2 \beta_2) \, .
\end{equation}

Note that, due to the $\pm$ sign in Eq.~\eqref{EQ:24}, if we have a pair of oscillatory eigenvalues with positive real part (growing with time), then the real part of the remaining two eigenvalues must necessarily be negative. It is interesting to note that oscillatory instabilities can only take place in the symmetry broken branches of the stationary solutions for any value of $B/A$; no oscillatory instability can be found on the symmetric branches of the stationary solutions where $\alpha = \alpha_1=\alpha_2$ and $\beta = \beta_1=\beta_2$, since, in this case, $S$ is always a real number. 

By evaluating partial derivatives with respect to the detunings and pump powers, we can also locate the point at which symmetry--breaking pitchfork bifurcations, corresponding to real eigenvalues becoming positive, occur. This critical point is given by

\vspace{-5mm}
\begin{equation}\label{eq:crit}
\frac{1 + \alpha_{1}^{2}}{2P_{1}\alpha_{1}} = \frac{1 + \alpha_{2}^{2}}{2P_{2}\alpha_{2}} = A - B\;. 
\end{equation}

\noindent This condition is the generalization of the critical point presented in Ref.~\cite{woodley2018universal} for $A=1$ and $B=2$.

Real eigenvalue instabilities can be found on the symmetric branches of the stationary solutions where $\alpha = \alpha_1=\alpha_2$ and $\beta = \beta_1=\beta_2$. Here, real $S$ means $S=2 |\gamma \alpha |$ and the conditions (\ref{EQ:22})-(\ref{EQ:23}) reduce to 

\vspace{-2.5mm}
\begin{eqnarray} \label{EQ:28}
& & |\gamma\alpha| > 1 + \alpha\beta\;, \\
 \label{EQ:29}
& & |\gamma\alpha| < -(1 + \alpha\beta)\;.
\end{eqnarray}

\noindent On the symmetric branches, the bifurcations corresponding to conditions \eqref{EQ:28} and \eqref{EQ:29} are either the saddle-node bifurcations of the S-shaped stationary curves or the pitchfork bifurcations leading to symmetry breaking solutions.

\begin{figure}[t]
\includegraphics[width=0.48\textwidth]{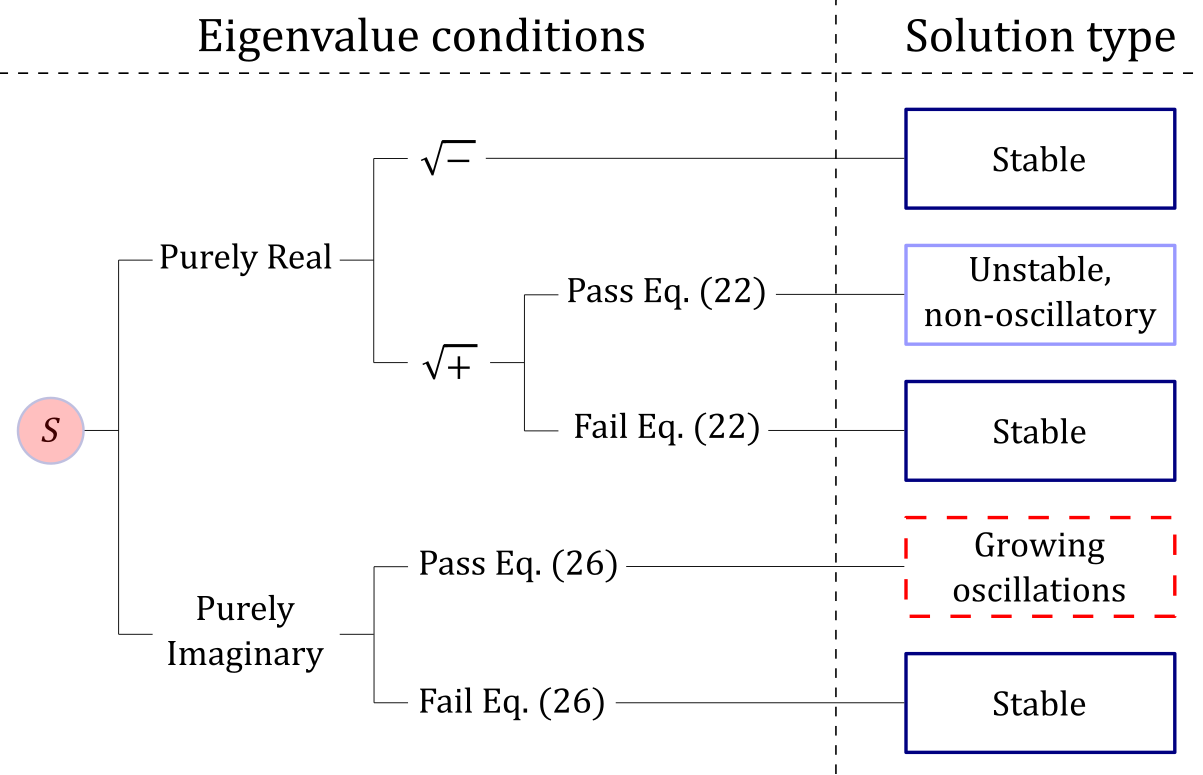}
\caption{Reference chart showing eigenvalue conditions, beginning from $S$, required to obtain solutions with various stability outcomes. This diagram should be used in reference to Section \ref{GenStab} for ease of understanding. Solution types are colour coded to match that of Figs. \ref{fig:7} and \ref{fig:8}.}
\label{fig:6}
\end{figure}

To illustrate the effect of the cross-phase modulation coefficient on the stability of the system, we report here about two limit cases of small and large cross-phase to self-phase modulation ratio $B/A$. We indicate the stable solutions with solid dark blue lines, the non-oscillatory instabilities with light blue lines, and the oscillatory instabilities with dashed red lines. Complex eigenvalues with positive real part may lead to oscillations that are experimentally accessible because their amplitude will eventually stop growing due to saturation effects that are not captured by the above linear stability analysis.

\begin{figure*}[t]\label{detuningPlots}
\includegraphics[width=\textwidth]{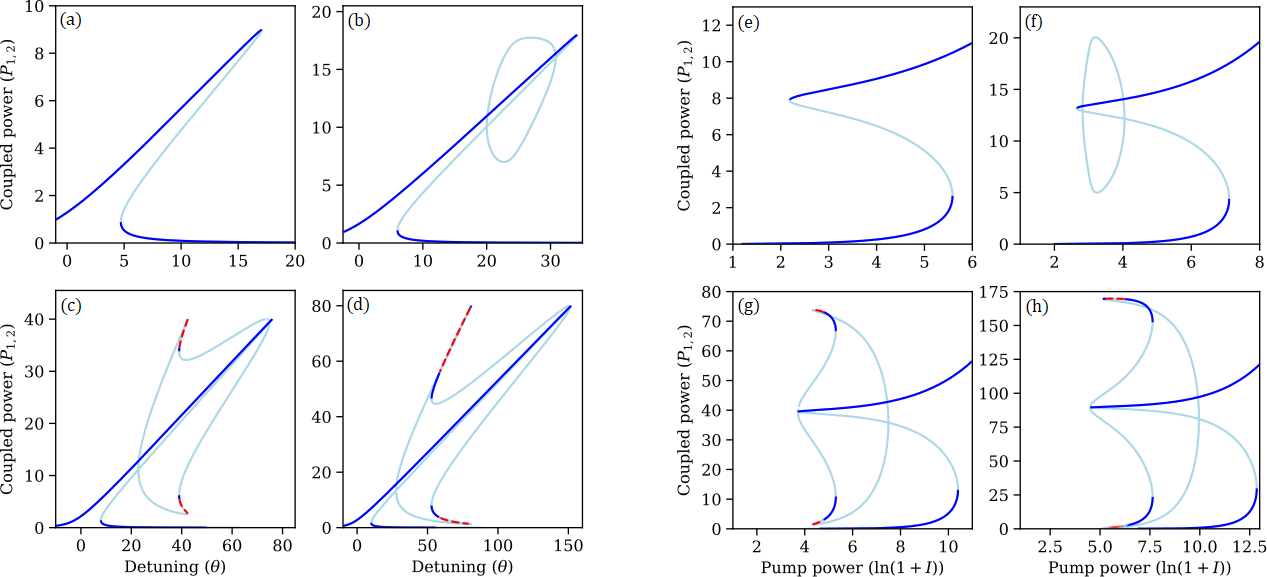}
\caption{(a)-(d) illustrate the coupled power $P_{1,2}$, against the detuning parameter, $\theta$ for $A=1$ and $B=0.9$. Stable and non-oscillatory unstable solutions are shown in dark and light blue, respectively, whilst oscillatory instabilities are shown in dashed red. The input power, $I$, increases with each frame. (a) $I=9$. No symmetry-broken solutions have yet emerged. (b) $I=18$. Spontaneous symmetry breaking occurs in the unstable symmetric branch. (c) $I=40$. Stable symmetry broken solutions emerge but quickly lose stability to complex eigenvalues with positive real parts. (d) $I=80$. The entire structure is stretched, including the region of unstable oscillatory and stable symmetry-broken solutions. Note the parameter range with four stable solutions. The right-hand plots illustrate the coupled power, $P_{1,2}$, against the input power, $I$. The key phenomena shown on the left are visible here. The advantage of this style of plot is that the relationship of the symmetry breaking and oscillations to the symmetric bistability (the S-shaped curve) can be seen. The detuning parameter, $\theta$, increases for each frame. (e) $\theta=15$. (f) $\theta=25$. (g) $\theta=75$. (h) $\theta=170$.  Note that, in the above plots, growing oscillations are always accompanied by a stable solution. Consequently, these oscillations may not be experimentally observable, since the system may instead favour the stable solution.}
\label{fig:7}
\end{figure*}

\begin{figure}
\includegraphics[width=0.455\textwidth]{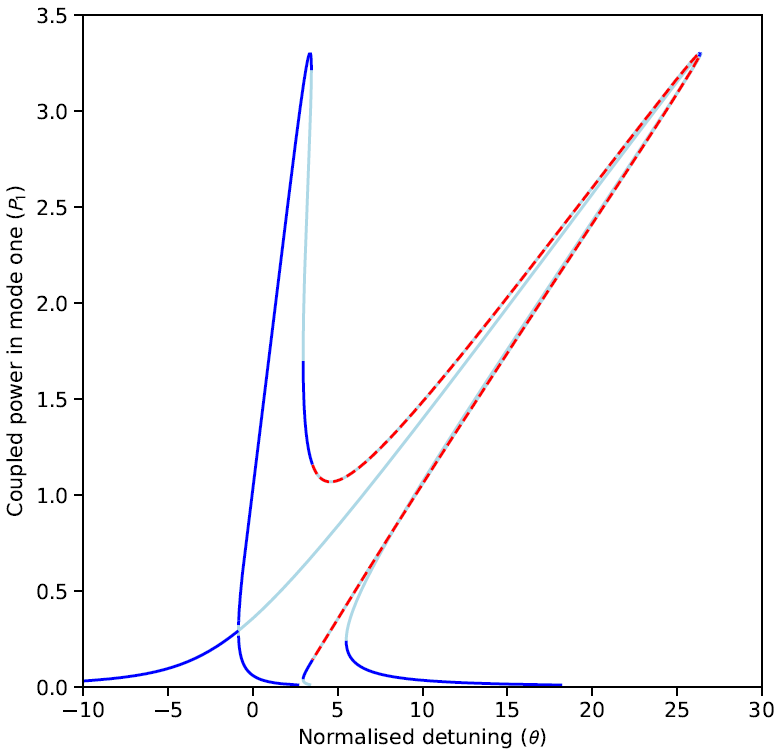}
\caption{Coupled power $P_{1,2}$, against the detuning parameter, $\theta$ for $A=1$, $B=7$ and $I=3.3$. Stable and non-oscillatory unstable solutions are shown in dark and light blue, respectively, whilst oscillatory instabilities are shown in dashed red. Note the large range of detunings over which growing oscillations exist.}
\label{fig:8}
\end{figure}

Figure \ref{fig:7} illustrates stable, unstable and oscillatory unstable regimes for a variety of choices of parameters for a small value of $B/A=0.9$, where the self-phase modulation is stronger than the cross-phase modulation. In this regime, the system is not strongly susceptible to either symmetry breaking or the onset of growing oscillations, and so the power thresholds for accessing these phenomena are very high. When increasing the input power, $I$, symmetry-broken solutions occur in the middle branch of the bistable S-shaped curves. Some of these solutions later gain stability, and others exhibit growing oscillations; the system in general begins displaying generalised multi-stability of symmetric and asymmetric solutions, as observed in Figs.~\ref{fig:7} (c), (d), (g), and (h).

For larger values of  $B/A$ such as $B/A=7$, large parameter regions where stationary states are susceptible to oscillations are observed, as displayed in the detuning scan in Fig. \ref{fig:8} for $I=3.3$. We expect widespread oscillatory regimes when the cross-phase modulation is larger than the self-phase modulation at experimentally attainable values of the pump power, $I$. Fig. \ref{fig:8} is also consistent with a prediction made in Section \ref{GenRes}: symmetry-broken solutions at zero detuning.

\section{Temporal Dynamics}

\begin{figure*}
\includegraphics[width=\textwidth]{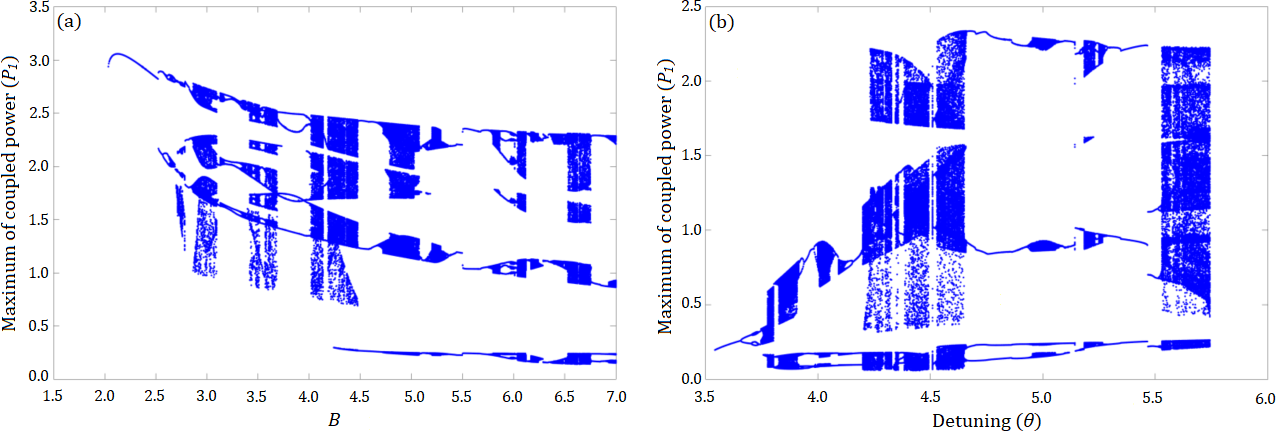}
\caption{(a) Poincar\'{e} sections of the maxima of oscillating coupled power $P_1$, versus the cross-phase modulation coefficient, $B$, for constant $A=1$. These points corresponds to constant values of the detuning $\theta=5$ and input power $I=3.3$. (b) Illustration of periodic oscillations and deterministic chaos in the Poincar\'{e} sections of the maxima of $P_{1}$ when varying the detuning parameter, $\theta$ for a large cross- to self-phase modulation ratio. In this case, $A=1$, $B=7$ and the input power is $I=3.3$. Note the dense columns of chaotic windows.}
\label{fig:9}
\end{figure*}

The previous section provides an important snapshot of the stability of the system -- specifically, how the system respond to small, noise like, perturbations  upon changes of the ratio $B/A$. In this section, we investigate the dynamics and possible oscillations by using numerical integration of Eq.~\eqref{EQ:19} for the full temporal evolutions. These numerical integrations illustrate the consequences of modifying the relative strengths of self- and cross-phase modulation for the onset of deterministic chaos and its extent in the system. We demonstrate here that increasing the value of $B$ increases the susceptibility of the system to temporal instability, and consequently chaos.

We firstly consider changes in the cross to self-phase modulation ratio $B/A$. For each parameter configuration specified by $B$, $\theta$, and $I$ in the oscillatory regime, we sample the evolution trajectories of the coupled powers $P_{1,2}$ by evaluating the Poincar\'{e} section corresponding to their local maxima where the first derivative in time is zero and second derivative is negative. In this way we can monitor the number of maxima per period and register their values. Fig. \ref{fig:9}(a) shows the maxima of the coupled power $P_1$ during oscillations when changing $B$ from 1.5 to 7, for $A=1$, $\theta=5$, and $I=3.3$. We observe sequences of bifurcations, chaotic windows and sudden crises. The power ranges spanned by the oscillations clearly increase with the cross-phase modulation magnitude.

To illustrate the susceptibility of the system to temporal oscillations at large values of $B/A$, we show in Fig.~\ref{fig:9}(b) the Poincar\'{e} sections in a detuning scan for $B/A=7$ and $I=3.3$. These are the same parameters of the stationary solution curves displayed in Fig.~\ref{fig:8}. In this case the symmetry breaking bifurcation occurs at negative values of the detuning $\theta$. After this bifurcation, one of the coupled powers increases while the other decreases. The onset of oscillations occurs when the decreasing coupled power approaches zero (just after $\theta=3.5$). Windows of periodic and chaotic oscillations alternate with increasing detunings until no symmetry broken solutions are observed just after $\theta=5.74$.

\begin{figure}[h!]\label{Test}
\includegraphics[width=\columnwidth]{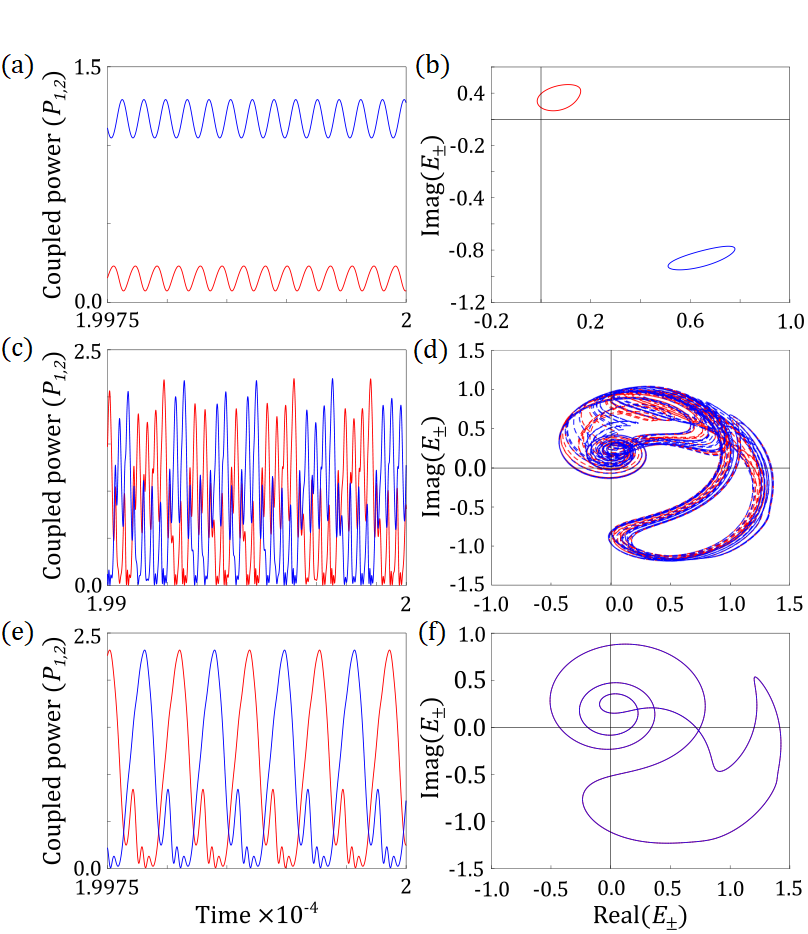}
\caption{Results of numerical simulations of Eq.\eqref{EQ:19} for $A=1$, $B=7$, and $I=3.3$ showing oscillatory (a), (b), chaotic switching (c), (d), and periodic switching (e), (f) solutions. (a), (c) and (e) all show the temporal evolutions of the coupled powers while (b), (d) and (e) show the phase space orbits of the real and imaginary components of the late temporal evolutions of $E_\pm$.  All simulations were run for $2\times 10^7$ iterations, with a time step of $dt=0.001$, starting from $E_+=0, E_-=0.0001$. The detuning parameter for (a), (b) is $\theta=3.6$, for (c), (d) is $\theta=4.31$ and for (e), (f) is $\theta=4.755$.}
\label{fig:10}
\end{figure}

The richness of oscillatory behaviour for $B/A=7$ and $I=3.3$ is presented in Fig.~\ref{fig:10}, which shows specific cases of different oscillatory regimes for given values of the detuning, as predicted by Fig.~\ref{fig:9}(b). Fig.~\ref{fig:10}(a),(b) show periodic oscillations close to the onset of temporal instability. Each asymmetrically coupled power has undergone a Hopf bifurcation, leading to a small amplitude modulation. The dynamical behaviour is attracted to two disjointed regions of the phase space. When increasing the detuning, the amplitude of the oscillations grows and chaotic dynamics are observed (see Fig.~\ref{fig:10}(c),(d)). We note, however, that the oscillations now switch erratically from one dominant field to the other and that the attractor covers a single region of the phase space for both coupled fields. This latter aspect becomes even more striking by a further increase in the detuning parameter as shown in Fig.~\ref{fig:10}(e),(f). Here, the system displays a periodic switching between the two modes and the projection of the attractors of the two fields overlap completely. An effect such as this has potential application in photonic systems where control of the output pulses, in particular of their polarization or propagation direction, is required. While we show this behaviour for $B/A=7$, we also predict that it would be present for many other values of the self- and cross- phase modulation constants.

The fact that the onset of chaos may be encouraged by increasing the relative strength of the cross-phase modulation (say, in the case of Kerr liquids, as compared to in a dielectric medium), is beneficial for potential applications of this chaotic regime -- for example, the realisation of all-optical polarisation scramblers.

\section{Conclusion}
We have presented a theoretical model for the spontaneous symmetry breaking of light in ring resonators, generalised to arbitrary strengths of self- and cross-phase modulation, and describing the coupling of either two circularly-polarised or two counter-propagating fields. We have presented the characteristics of the steady-state symmetry-broken region, such as the minimum criteria for its observation, its opening and closing bifurcation points and the conditions for maximum difference in the coupled intensities. It was observed how the position of the symmetry-broken region varies with respect to the symmetric optical bistability, along with the dependence of the oscillatory regime on the value of $B/A$. Finally, we have shown the possible presence of a subset of oscillatory solutions which may lead to new applications such as sequences of pulses with given polarization or propagation direction. These oscillatory behaviours include different styles of (chaotic and periodic) switching between modes. Periodic switching suggests a transition to self-organising behaviour in a chaotic regime. These findings should be applicable to a large range of experimental settings featuring nonlinear media, including Kerr liquids and atomic vapours, as well as situations that exhibit variable overlap (and, hence, variable cross-phase modulation) between two optical modes.
\\
\section{Acknowledgements}
We acknowledge financial support from: EPSRC DTA Grant No. EP/M506643/1; H2020 Marie
Sklodowska-Curie Actions (MSCA) (748519, CoLiDR); National Physical Laboratory Strategic Research; H2020 European Research Council (ERC) (756966, CounterLight); Engineering and Physical Sciences Research Council (EPSRC).

The authors would very much like to thank Jonathan M. Silver and Leonardo Del Bino for useful and stimulating discussions.

\
\bibliographystyle{ieeetr}
\bibliography{Ref}

\end{document}